\documentclass[aps,prb,preprint,groupedaddress]{revtex4-1}
\usepackage{graphicx}
\usepackage{float}
\usepackage{amsmath}
\DeclareMathOperator\erf{erf}
\DeclareMathOperator\erfc{erfc}
\begin{document}


\title{Real space pairwise electrostatic summation in a uniform neutralising background}

\author{Chris J.\ Pickard}
\email[]{cjp20@cam.ac.uk}
\affiliation{Department of Materials Science \& Metallurgy, University of Cambridge, 27 Charles Babbage Road, Cambridge CB3~0FS, United Kingdom}
\affiliation{Advanced Institute for Materials Research, Tohoku University 2-1-1 Katahira, Aoba, Sendai, 980-8577, Japan}


\date{\today}

\begin{abstract}
Evaluating the total energy of an extended distribution of point charges, which interact through the Coulomb potential, is central to the study of condensed matter. With near ubiquity, the summation required is carried out using Ewald's method, which splits the problem into two separately convergent sums; one in real space and the other in reciprocal space.  Density functional based electronic structure methods require the evaluation of the ion-ion repulsive energy, neutralised by a uniform background charge. Here a purely real-space approach is described. It is straightforward to implement, computationally efficient and offers linear scaling. When applied to the evaluation of the electrostatic energy of neutral ionic crystals, it is shown to be closely related to Wolf's method.
\end{abstract}


\maketitle

\section{Introduction}

The total energy of a collection of $N$ point charges is given by:
\begin{equation}
E=\frac{1}{2}\sum_i^{N}\sum_{j\ne i}^{N}\frac{Z_iZ_j}{|{\bf r}_i-{\bf r}_j|}=\frac{1}{2}\sum_i^{N}\sum_{j\ne i}^{N}\frac{Z_iZ_j}{r_{ij}},
\end{equation}
where $Z_i$ is the electrostatic charge on ion $i$, located at ${\bf r}_i$, and $r_{ij}$ is the distance between ion $i$ and ion $j$. Hartree atomic units are used here and throughout. The study of condensed matter demands the treatment of extended systems, where $N$ is very large (and frequently taken to be infinite). Extended systems are typically handled through the imposition of periodic boundary conditions, in which periodic replicas of a small part of the system are repeated through space.  This is a natural approach for perfect crystals, and non-crystalline systems can also be accommodated through the use of supercells. The electrostatic energy per unit cell, $E_{\rm cell}$, can be written within periodic boundary conditions as:
\begin{equation}
E_{\rm cell}=\frac{1}{2}\sum_i^{N_{\rm cell}}\sum_{j\ne i}^{\infty}\frac{Z_iZ_j}{r_{ij}}=\sum_i E_i,
\end{equation}
where $N_{\rm cell}$ is the number of ions in a unit cell. The remainder of this article will focus on the evaluation of the electrostatic energy for a single ion, $i$, interacting with all the others:
\begin{equation}
E_i=\frac{1}{2}\sum_{j\ne i}^{\infty}\frac{Z_iZ_j}{r_{ij}}.
\label{ei}
\end{equation}

Following the introduction by Wolf and co-workers\cite{wolf1999exact} of an accurate and efficient alternative to Ewald summation\cite{ewald1921berechnung,de1980simulationI,de1980simulationII} there has been considerable interest in so-called non-Ewald methods.\cite{zahn2002enhancement,wu2005isotropic,elvira2014damped,lamichhane2014real,fanourgakis2015extension,muscatello2011comparison,ma2005modified,ojeda2015treating,fukuda2012non,gdoutos2010comparison} Ewald's method is based on partitioning the sum in Eqn. \ref{ei} into two parts by scaling the Coulomb interaction by the sum of the error and complementary error functions (recalling that $\erf(x)+\erfc(x)=1$):
\begin{equation}
E_i=\frac{1}{2}\sum_{j\ne i}^{\infty}\frac{Z_iZ_j}{r_{ij}}\erfc\left(\frac{r_{ij}}{R_d}\right)+\frac{1}{2}\sum_{j\ne i}^{\infty}\frac{Z_iZ_j}{r_{ij}}\erf\left(\frac{r_{ij}}{R_d}\right).
\end{equation}
The rapid decay of the complementary error function with $r_{ij}$ allows the first part of the summation to be straightforwardly converged, including only those interactions within the locality of ion $i$. The second term is evaluated in reciprocal space, where it also converges rapidly. The resulting algorithm is a mainstay of computational physics and chemistry. While mathematically and computationally elegant, Ewald's method does not lend itself straightforwardly to a real space physical interpretation. Motivated by this, Wolf and co-workers took a fresh look\cite{wolf1999exact} at the problem and analysed the convergence of the real space lattice sum in the evaluation of the Madelung energy for ionic crystals. It was concluded that much of the observed poor convergence could be attributed to the oscillating violation of charge neutrality within the real space cutoff sphere. As previously proposed by Adams,\cite{adams1979computer} this was rectified by placing the charge deficit on the surface of the cutoff sphere. When combined with the damping of the real space term (as in Ewald's method) this led to a rapidly convergent, computationally simple, and efficient alternative to the dual space approach of Ewald.\cite{} Some have asked whether Ewald's summation is still necessary.\cite{fennell2006ewald}  

An important area of computational physics and chemistry that still very much depends on the original Ewald scheme is density functional theory (DFT)\cite{hohenberg1964inhomogeneous,kohn1965self} based total energy electronic structure methods.\cite{martin2004electronic,payne1992iterative} These methods have been taken to constitute a ``standard model'' for the materials sciences.\cite{hasnip2014density} High quality and benchmarked implementations\cite{lejaeghere2016reproducibility} are extremely widely used to calculate materials properties, interpret and complement experiments, and even predict new crystal structures and their defects.\cite{pickard2011ab} The DFT total energy can be written as:
\begin{equation}
\begin{aligned}
E_{\rm tot}[\rho,\{{\bf r}_i\}]=T[\rho]+E_{\rm eN}[\rho,\{{\bf r}_i\}]+E_{\rm H}[\rho]+E_{\rm xc}[\rho]+E_{\rm NN}[\{{\bf r}_i\}].
\end{aligned}
\label{DFT}
\end{equation}
Leaving aside the kinetic energy ($T[\rho]$) and exchange correlation ($E_{\rm xc}[\rho]$) terms, the external potential ($E_{\rm eN}[\rho,\{{\bf r}_i\}]$), Hartree ($E_{\rm H}[\rho]$) and nucleus-nucleus (or ion-ion) electrostatic ($E_{\rm NN}[\{{\bf r}_i\}]$) terms are not individually defined in an extended system. However, for an overall charge neutral system (the total number of electrons being equal to the total charge of the ions) their sum is. Given that they are evaluated separately for computational reasons, these individual terms are tamed by inserting uniform neutralising background charges. It is for this reason that Wolf's scheme, and its derivatives, are not suitable for the evaluation of $E_{\rm NN}[\{{\bf r}_i\}]$. The placing of the entire neutralising charge on to the surface of the cutoff sphere is physically incorrect. It should be spread through space.

In this article a real space summation approach, suitable for application to density functional electronic structure methods for extended systems, will be described. It is based on ion/nucleus centred neutralising \emph{spheres} of charge. The choice of the radius of these spheres is shown to be critical to the success of the method. Rapid convergence with real space cutoff is assured by damping the Coulomb interaction, and analytically correcting the errors introduced by the damping. The approach can also be applied to the evaluation of the Madelung energy of ionic crystals, and in this special case it is shown to be closely related to the Wolf method. Instead of the compensating charge being placed on the surface of the cutoff sphere, it is distributed throughout a shell, which has a finite thickness.

\section{Real space cutoff}

\begin{figure}[]
\centering
\includegraphics[width=0.475\textwidth]{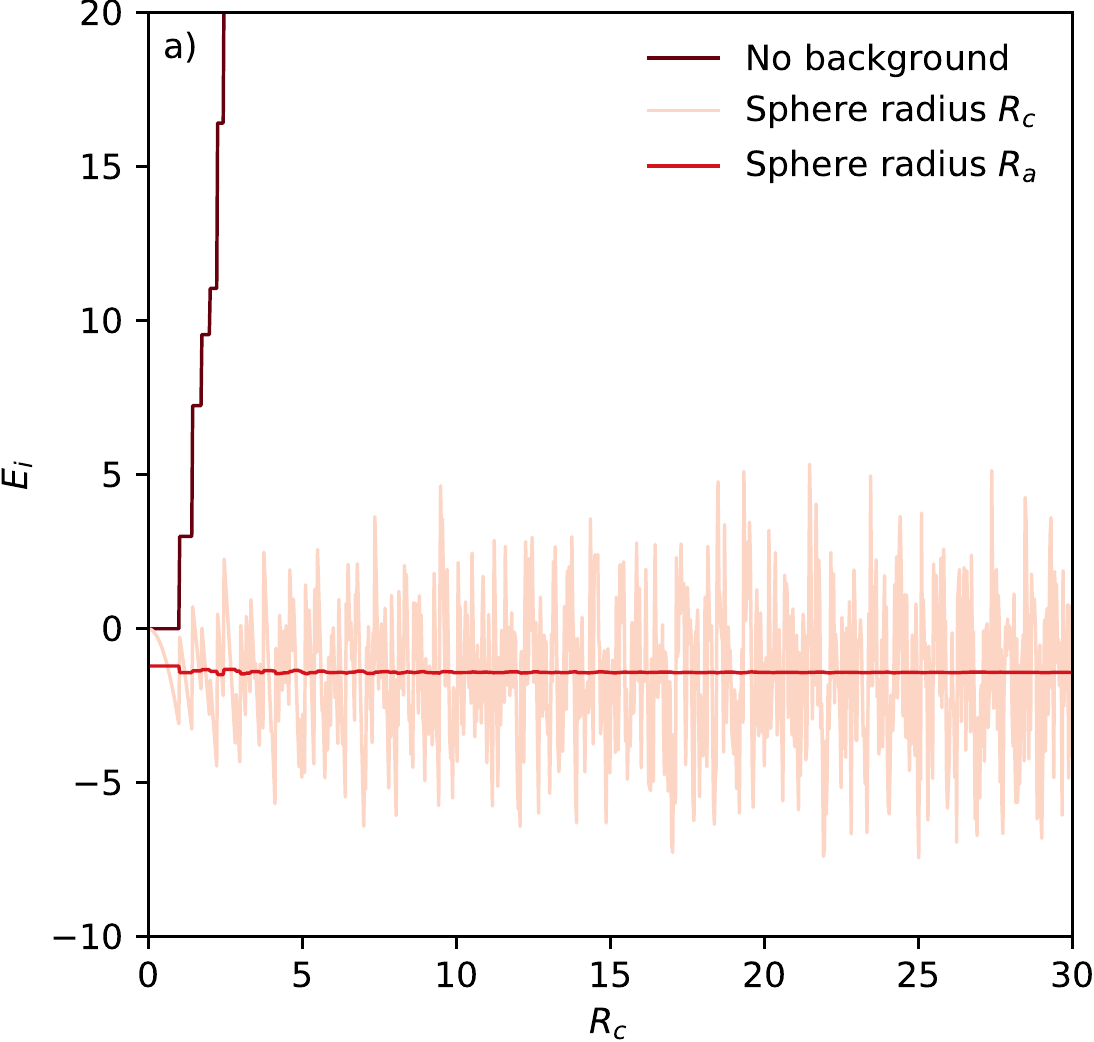}
\includegraphics[width=0.49\textwidth]{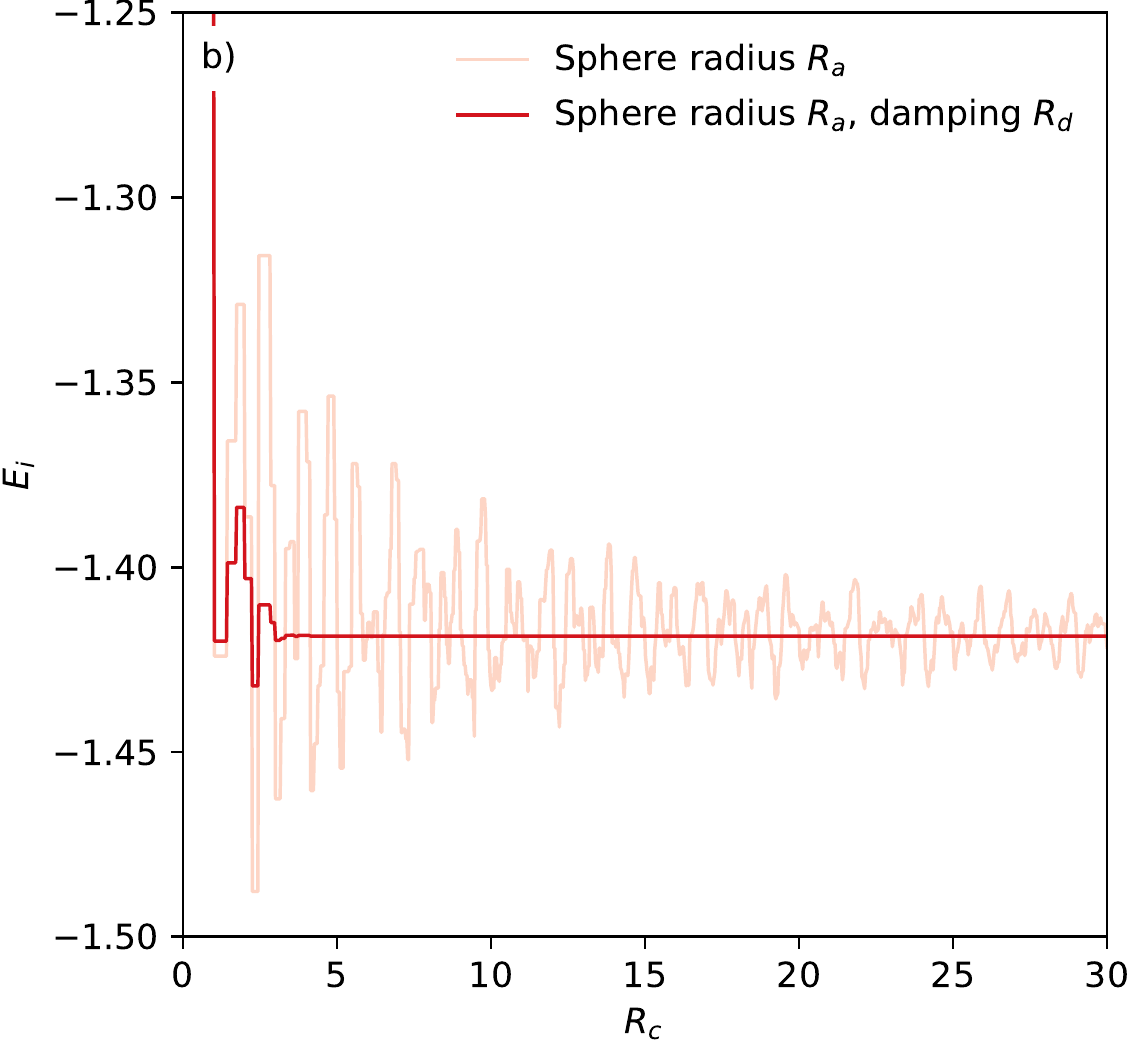}
\caption{Real space pairwise electrostatic sum for a simple cubic lattice and nearest neighbour distance of $1$. a) With no neutralising background the sum diverges with increasing $R_c$. Choosing a neutralising background sphere of radius, $R_c$, the sum stop increasing with $R_c$, but it does not converge. In contrast, an adaptive radius, $R_a$, ensures convergence. b) Damping rapidly accelerates convergence ($R_d=1.5$).}\label{figure1}
\end{figure}

The first step in any practical scheme is to restrict the summation over $j$ in Eqn. \ref{ei}. In a \emph{homogeneous} system, for example a crystal, the obvious approach is to define a sphere (with a radius of $R_c$, and centred on atom $i$) beyond which contributions to $E_i$ are neglected. The sum then becomes:
\begin{equation}
E_i=\frac{1}{2}\sum_{j\ne i,r_{ij}<R_c}^{\infty}\frac{Z_iZ_j}{r_{ij}}.
\label{eicut}
\end{equation}
Given the long ranged nature of the Coulomb interaction, when all the point charges are of the same sign, this sum can only grow with $R_c$. This rapid increase is demonstrated in Fig. \ref{figure1} for a simple cubic lattice, and the growth approximately follows the square of the cutoff radius, $R_c$.

\section{Charge neutralisation}

As discussed in the introduction, the ion-ion electrostatic interaction energy, $E_{\rm NN}$, required in DFT total energy calculations, is to be computed in the presence of a neutralising uniform background charge. For a crystal, or a system modelled by a supercell, the average charge density due to the ions, $\rho$, is given by:
 \begin{equation}
\rho=\sum_i^{N_{cell}}Z_i/\Omega=Q/\Omega,
\end{equation}
where $\Omega$ is the volume of the unit cell and $Q$ is the total charge of the ions in the cell. Integrating over the uniform neutralising background charge, which is given by $-\rho$, and cutting the integral off in the same way as the sum the energy for ion $i$, $E_i$, is given by:
\begin{equation}
\begin{aligned}
E_i&=\frac{1}{2}\sum_{j\ne i,r_{ij}<R_c}^{\infty}\frac{Z_iZ_j}{r_{ij}}-\frac{1}{2}\int_{r<R_c}\frac{Z_i\rho}{r} d^3r \\
&=\frac{1}{2}\sum_{j\ne i,r_{ij}<R_c}^{\infty}\frac{Z_iZ_j}{r_{ij}}-\pi Z_i\rho R_c^2.
\end{aligned}
\end{equation}
In Fig. \ref{figure1}, the rapid growth of $E_i$ can be seen to have been eliminated. However, there is no meaningful convergence, as a result of the imperfect neutralisation of the charge within the cutoff sphere for a general value of $R_c$.

\section{Adaptive cutoff radius}

For a given value of $R_c$ the total charge of the included ions can be evaluated:
\begin{equation}
Q_i=\sum_{j,r_{ij}<R_c}^{\infty}Z_j.
\end{equation}
In general $Q_i\neq\frac{4\pi}{3}R_c^3\rho$, and the compensated spherically truncated system is left with an overall charge. To enforce charge neutrality, an adaptive radius for the compensation charge sphere can be chosen:
\begin{equation}
Q_i=\frac{4\pi}{3}R_a^3\rho \implies R_a=\sqrt[3]{\frac{3Q_i}{4\pi\rho}}.
\label{Qadapt}
\end{equation}
Using this radius for the compensating sphere has a dramatic impact, in that:
\begin{equation}
E_i=\frac{1}{2}\sum_{j\ne i,r_{ij}<R_c}^{\infty}\frac{Z_iZ_j}{r_{ij}}-\pi Z_i\rho R_a^2,
\label{adapt}
\end{equation}
converges with increasing $R_c$, as demonstrated in Fig.\ref{figure1}. The convergence with $R_c$ is, however, oscillatory and slow, and the computational scheme is not yet useful.

\section{Damping}

The oscillatory convergence of the adaptive cutoff scheme can be explained by the discrete inclusion of ions as the cutoff sphere expands. It can be eliminated by smoothly reducing (or damping) the contribution of the more distant ions to the real space sum. Changing the terms in the real space sum alters the total and an analytic correction to this damping must be applied to minimise the error introduced. In the following the damping procedure will be described.

As in the Ewald method, each ion within the cutoff sphere is dressed with a neutralising spherical Gaussian charge distribution, containing an equal but opposite charge, and an extent that is controlled by $R_d$,
\begin{equation}
\rho_d^j(r)=\frac{-Z_j}{\pi^{3/2}R_d^3}e^{-r^2/R_d^2}.
\label{rhod}
\end{equation}
Evaluating the combined electrostatic potential due to the charge distributions $\rho_d^j(r)$ and the point charges $Z_j$, the energy for ion $i$, $E_i$, is given by:
\begin{equation}
\begin{aligned}
E_i=&\frac{1}{2}\sum_{j\ne i,r_{ij}<R_c}^{\infty}\frac{Z_iZ_j\erfc(r_{ij}/R_d)}{r_{ij}} +\Delta E_i.
\end{aligned}
\label{dmpcut}
\end{equation}
The $\Delta E_i$ term in the above is made up of three parts:

\begin{equation}
\begin{aligned}
\Delta E_i=\Delta E_{\rm sphere}+\Delta E_{\rm damp}+\Delta E_{\rm self}.
\end{aligned}
\label{deltaei-parts}
\end{equation}
The first term, $\Delta E_{\rm sphere}$, describes the interaction of $Z_i$ with the uniform compensating sphere (radius $R_a$) of charge with a density of $-\rho$. It is identical to that in Eqn.\ref{adapt}: 

\begin{equation}
\begin{aligned}
\Delta E_{\rm sphere}=-\frac{Z_i}{2}\int_0^{R_a}\frac{\rho}{r}d^3r=-\pi Z_i\rho R_a^2.
\end{aligned}
\end{equation}

The second term, $\Delta E_{\rm damp}$, provides a correction to the error introduced by the damping.  This correction is constructed by taking the charge associated with the ions to be distributed uniformly throughout the cutoff sphere, centred on ion $i$, with a density $-\rho$ (negative so as to cancel the dressing terms in the sum). This uniform charge distribution is dressed by performing a convolution with the charge dressing function $\rho_d$ (per unit charge, i.e. setting $Z_j=1$ in Eqn. \ref{rhod}). The resulting convolved charge distribution is spherically symmetric, by construction it is independent of the detailed positions of the ions within the cutoff sphere, and it approximately cancels the dressing charges centred on the ions:
\begin{equation}
\begin{aligned}
\rho_a(r)=\frac{\rho}{2r}\bigg(\frac{R_d(e^{-\frac{(r+R_a)^2}{R_d^2}}-e^{-\frac{(r-R_a)^2}{R_d^2}})}{\sqrt{\pi}}+r\erf(\frac{r+R_a}{R_d})-r\erf(\frac{r-R_a}{R_d})\bigg).
\end{aligned}
\end{equation}
The damping correction, $\Delta E_{\rm damp}$, can now be written as the interaction of this charge distribution with the central ion $i$:
\begin{equation}
\begin{aligned}
\Delta E_{\rm damp}=\frac{Z_i}{2}\int_0^{\infty}\frac{\rho_a(r)}{r}d^3r=\pi Z_i\rho (R_a^2-R_d^2/2)\erf(R_a/R_d)+\sqrt{\pi} Z_i\rho R_aR_de^{-R_a^2/R_d^2}.
\end{aligned}
\end{equation}
The final term, $\Delta E_{\rm self}$, describes the interaction of $Z_i$ with the Gaussian charge distribution dressing it (the self term, which is not included in the sum):
\begin{equation}
\begin{aligned}
\Delta E_{\rm self}=\frac{Z_i}{2}\int_0^{\infty}\frac{\rho_d}{r}d^3r=-\frac{1}{\sqrt{\pi}R_d}Z_i^2.
\end{aligned}
\end{equation}
The final expression for $\Delta E_i$ is given by: 
\begin{equation}
\begin{aligned}
\Delta E_i=-\pi Z_i\rho R_a^2+\pi Z_i\rho (R_a^2-R_d^2/2)\erf(R_a/R_d)+\sqrt{\pi} Z_i\rho R_aR_de^{-R_a^2/R_d^2}-\frac{1}{\sqrt{\pi}R_d}Z_i^2.
\end{aligned}
\label{deltaei}
\end{equation}
When $R_c\gg R_d$ (and hence $R_a\gg R_d$), the above simplifies to:
\begin{equation}
\begin{aligned}
\Delta E_i=-\pi Z_i\rho R_d^2/2-\frac{1}{\sqrt{\pi}R_d}Z_i^2.
\end{aligned}
\end{equation}
As demonstrated in Fig. \ref{figure1}, the damped scheme (using Eqns. \ref{dmpcut} and \ref{deltaei}) exhibits rapid convergence with $R_c$ when $R_a$ is evaluated according to Eqn. \ref{Qadapt}.

\section{Parameter choice}

\begin{figure}[]
\centering
\includegraphics[width=0.475\textwidth]{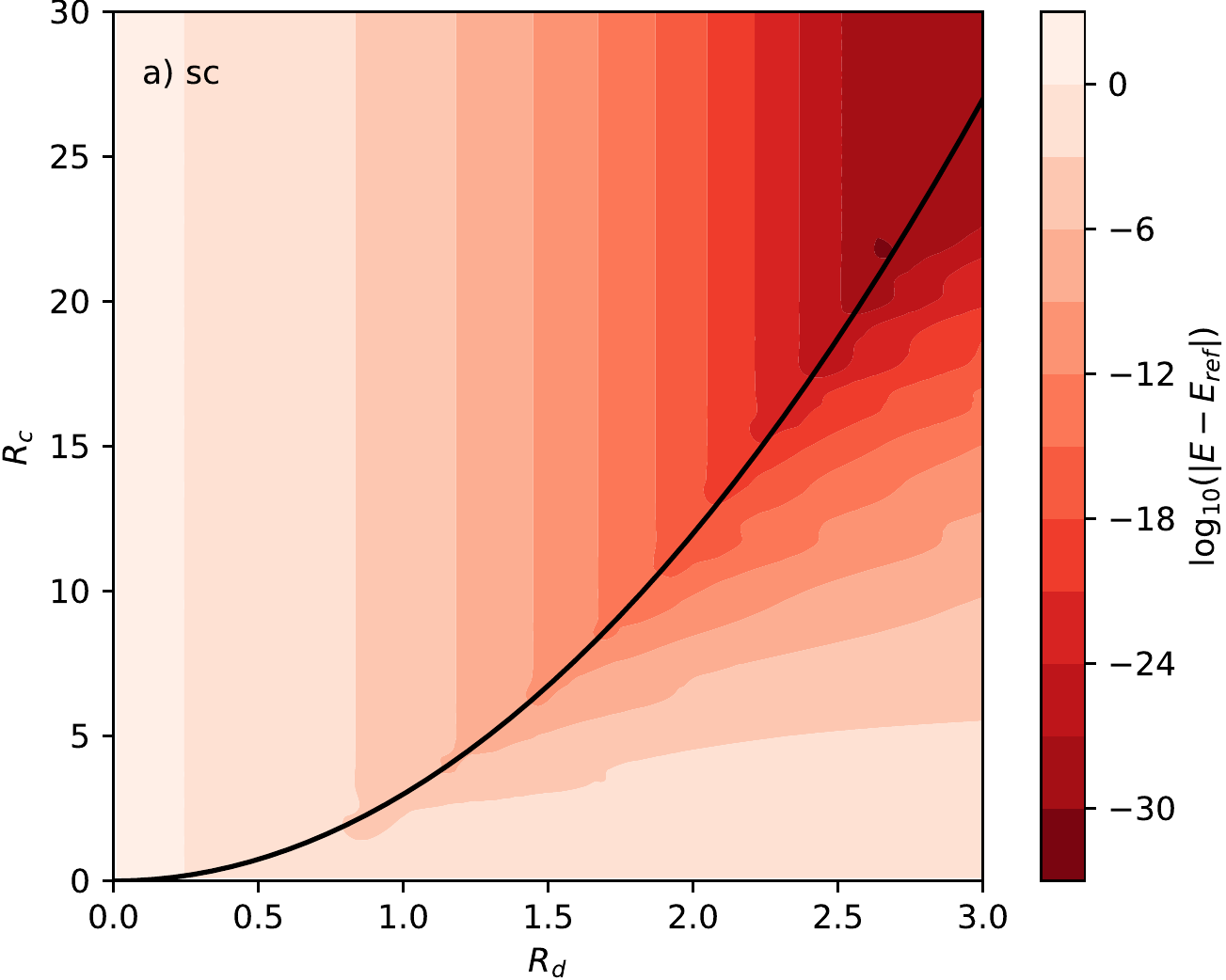}
\includegraphics[width=0.475\textwidth]{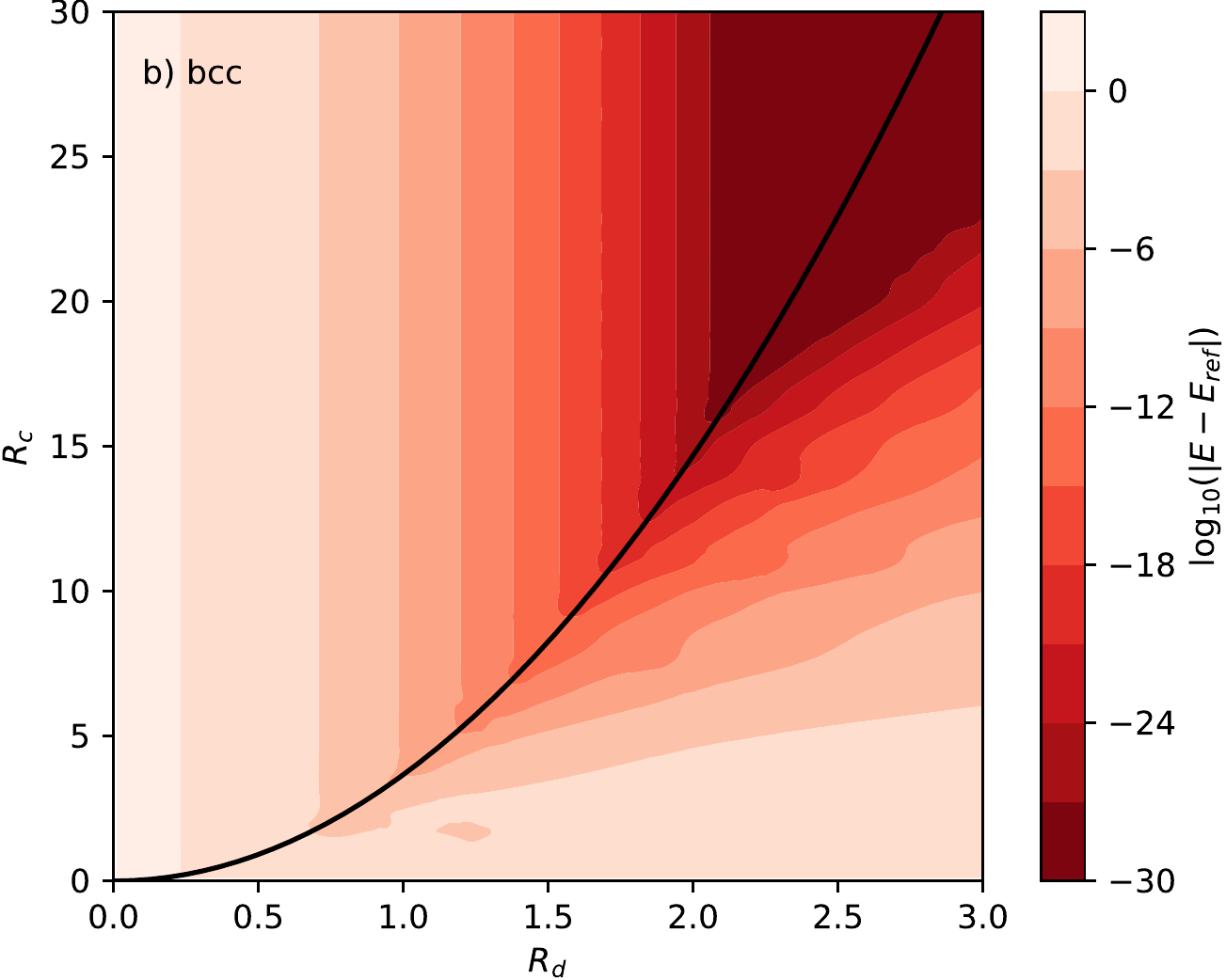}
\includegraphics[width=0.475\textwidth]{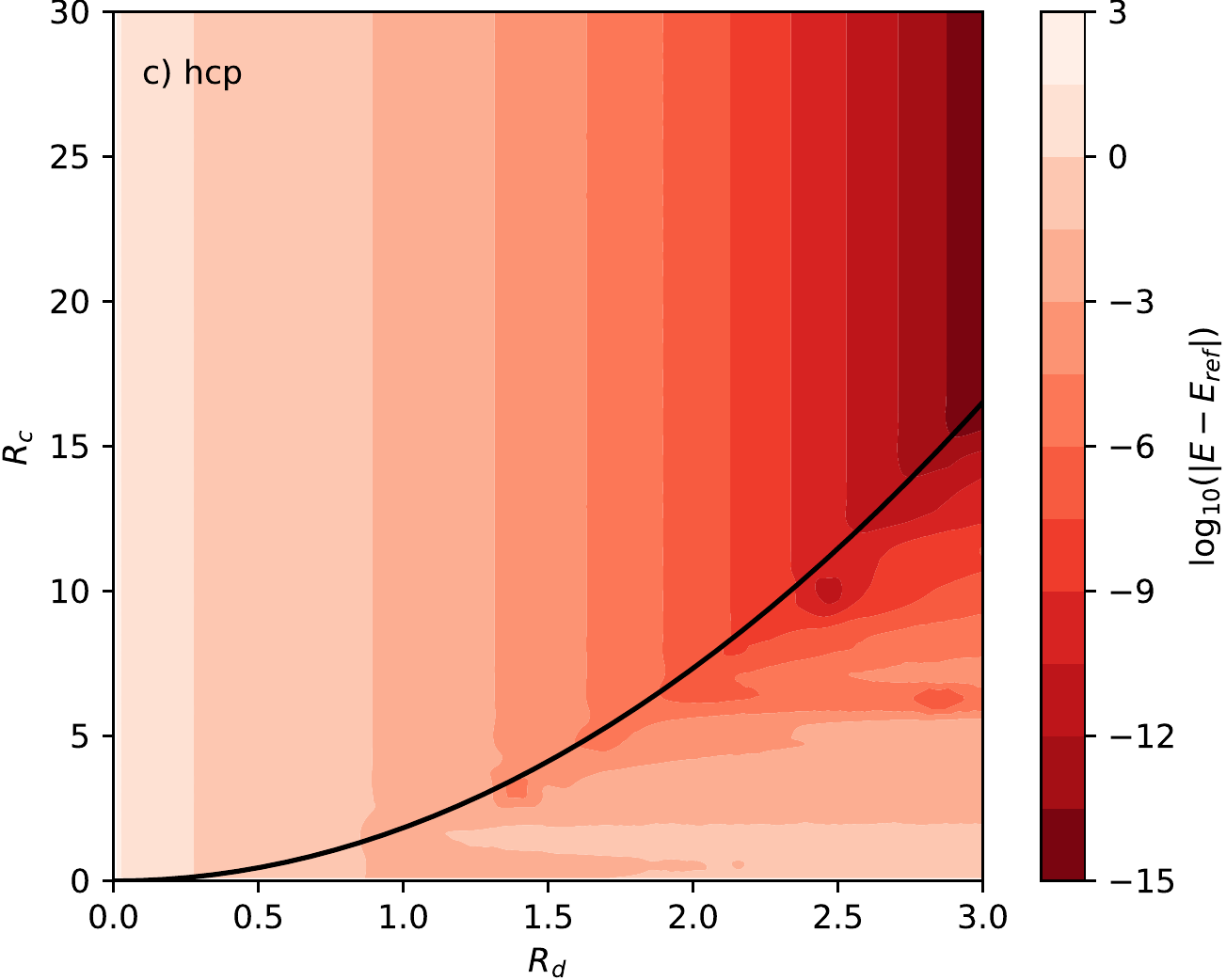}
\includegraphics[width=0.475\textwidth]{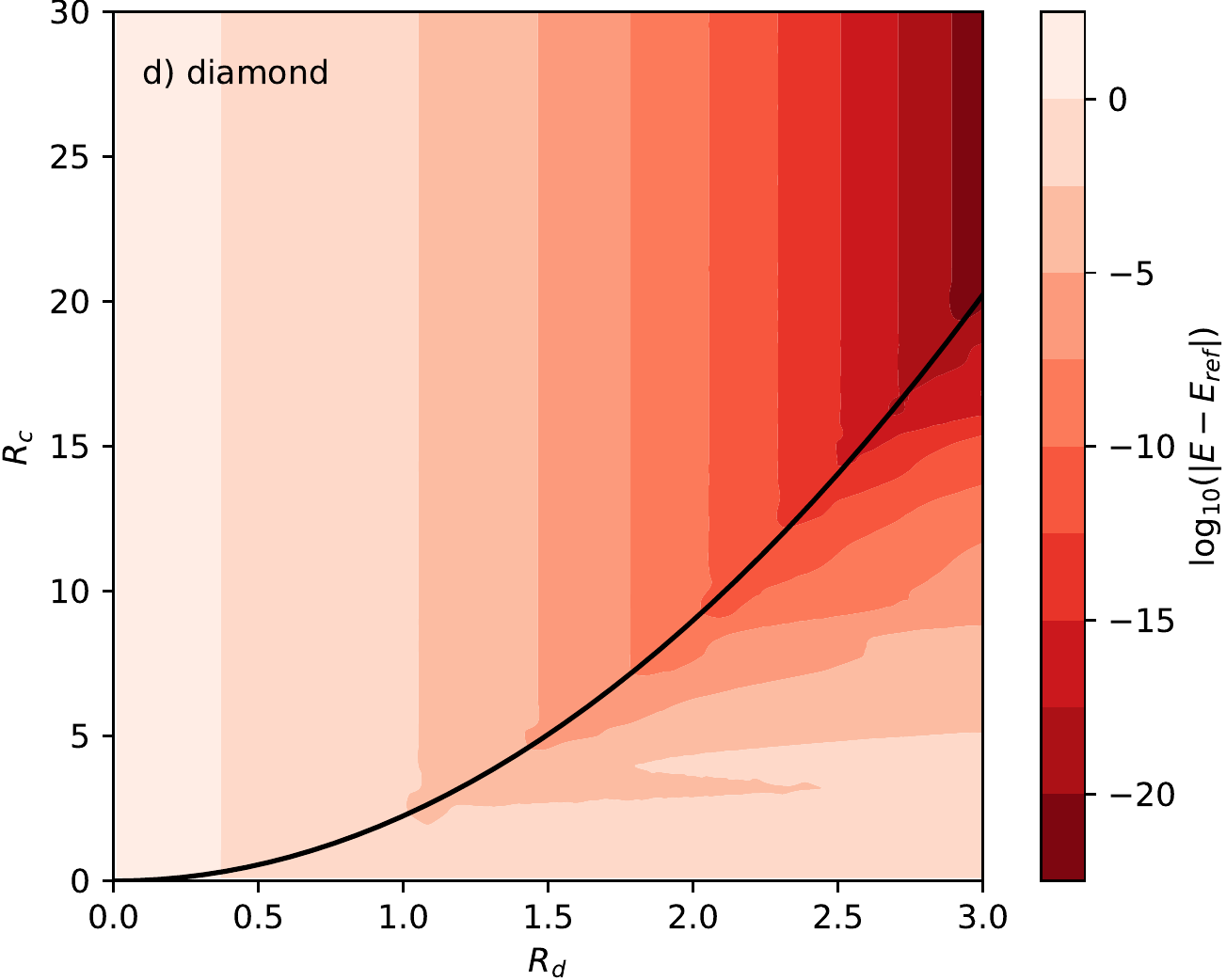}
\caption{Logarithmic error in the real space pairwise damped sum, for a) simple cubic ($h_{\rm max}=1$), b) body centred cubic ($h_{\rm max}=\sqrt{2/3}$), c) hexagonal close packed ($h_{\rm max}=\sqrt{8/3}$), and d) diamond ($h_{\rm max}=4/3$) lattices. The nearest neighbour distance is $1$ for all lattices. The reference energy, $E_{\rm ref}$, is evaluated for $R_c=36$ and $R_d=4$. The black line indicates an optimal path to convergence, and provides a relationship between $R_c$ and $R_d$: $R_c=3R_d^2/h_{\rm max}$.}
\label{figure2}
\end{figure}

Two parameters have been introduced: the radius of the cutoff sphere ($R_c$), and the damping parameter ($R_d$). While this is one fewer than for the Ewald scheme (no reciprocal space cutoff is required) it would be considerably more convenient if there were just a single parameter, which directly controlled the accuracy of the final result. It is apparent that there should be some relationship between an ideal choice of $R_c$ and $R_d$. For fixed $R_d$, the contribution from distant ions rapidly diminishes, and with increasing $R_c$ the correction term quickly converges. [Note that $\erfc(10)=2.1\times10^{-45}$] And so, increasing $R_c$ (with the associated computational cost) beyond a few multiples of $R_d$ will not lead to a more accurate result (the remaining error being unavoidable, and due to the uncorrectable effects of the damping). An improvement in accuracy can only be achieved by increasing $R_d$ (i.e. decreasing the damping) with $R_c$. In Fig. \ref{figure2} the convergence of $E_i$, towards numerical results established from a high cutoff and minimally damped calculations, is investigated. For the lattices considered, a relationship between $R_c$ and $R_d$ emerges: $\hat R_c=3\hat R_d^2$, where $\hat R_c=R_c/h_{\rm max}$ and $\hat R_d=R_d/h_{\rm max}$. The length scale, $h_{\rm max}$, can be calculated from the lattice as the largest perpendicular distance between the faces of the primitive cell, or chosen to represent known length-scales or features of the system.

In Table \ref{table1} the adaptively cutoff, and damped, method is benchmarked against an implementation of Ewald's scheme incorporated in the CASTEP code\cite{clark2005first} (version 18.1, using default settings). Taking $\hat R_d=2$ and $\hat R_c=3\hat R_d^2$, agreement to 9 or 10 significant figures is readily achieved.
\begin{table}[]
\centering
\caption{$E_{\rm NN}$ calculated using the current scheme, and the Ewald summation implemented in CASTEP\cite{clark2005first}. The valence charges used are: $Z_{\rm Al}=+3$, $Z_{\rm Si}=+4$, $Z_{\rm O}=+6$}
\label{table1}
\begin{tabular}{l|l|l|l|l|l|r|r}
\hline
Composition & fu & Space Group  & ICSD\cite{hellenbrandt2004inorganic} coll. code& $h_{\rm max}$&$\hat R_d$ & $E_{NN}$ (Current) & $E_{NN}$ (Ewald) \\ \hline
Al                  & 1 & Fm$\bar 3$m &     43423          &    4.42         & 2.0&  -2.695954572 & -2.695954572         \\
                     &   &                        &                          &                    & 1.5&  -2.695954572 &           \\
                     &   &                        &                          &                    & 1.0&  -2.696016437 &           \\\hline
Si                  &  2 & Fd$\bar 3$m &    51688            &   5.92         & 2.0&  -8.398574646  & -8.398574646       \\
                     &   &                        &                          &                    & 1.5&  -8.398574646 &           \\
                     &   &                        &                          &                    & 1.0&  -8.398667787 &           \\\hline
SiO$_2$         &  3 & P3$_1$21     &   29122             &    10.21       & 2.0&  -69.488098659 & -69.488098658           \\
                     &   &                        &                          &                    & 1.5&  -69.488098654 &           \\
                     &   &                        &                          &                    & 1.0&  -69.487429611 &           \\\hline
Al$_2$SiO$_5$ &  4 &     Pnnm    &    24275             &   14.93        & 2.0&  -244.055008450 & -244.055008300            \\ 
                     &   &                        &                          &                    & 1.5&  -244.055008299 &           \\
                     &   &                        &                          &                    & 1.0&  -244.054904540&           \\\hline
\end{tabular}
\end{table}

\begin{figure}[]
\centering
\includegraphics[width=0.95\textwidth]{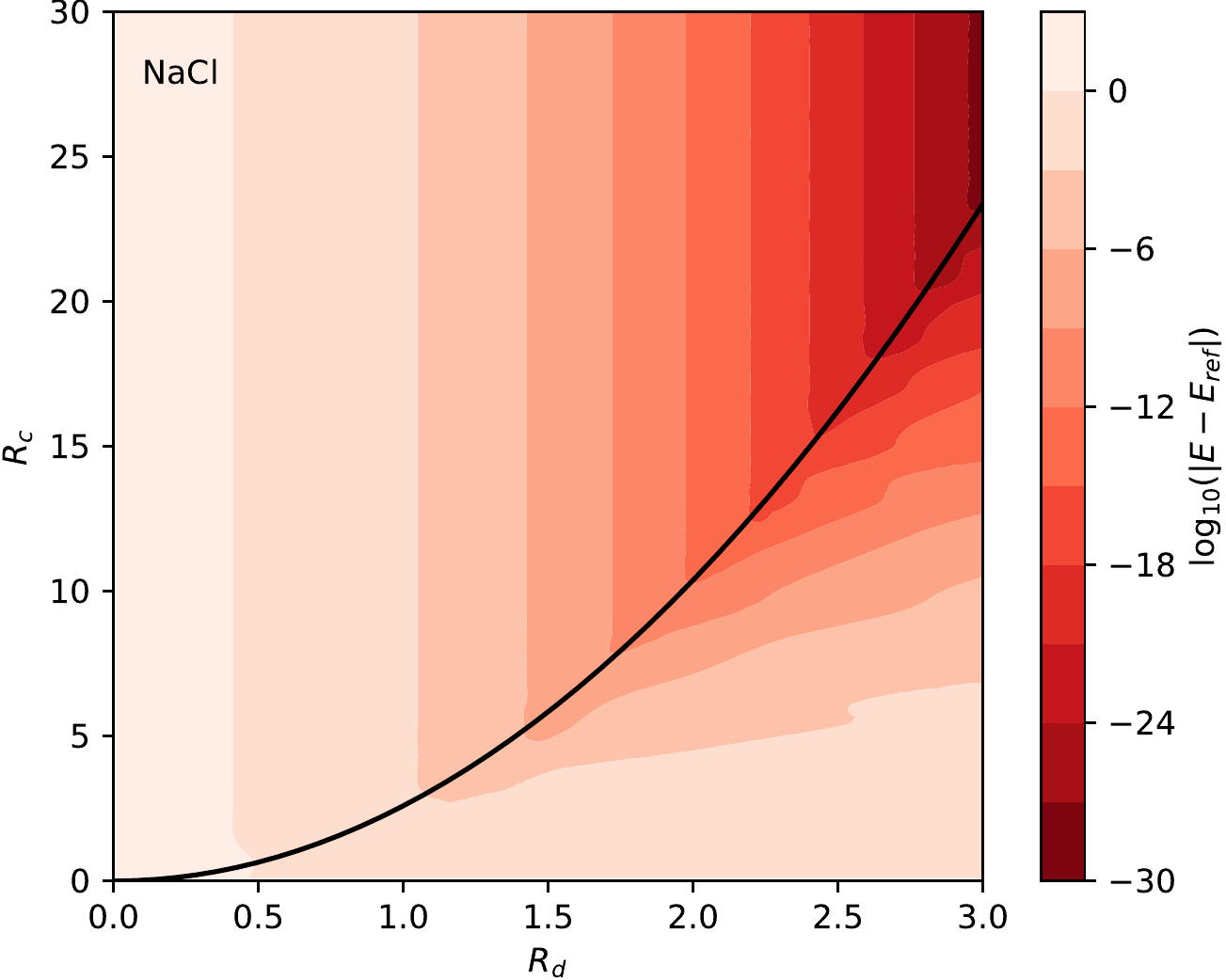}
\caption{Logarithmic error in the real space pairwise damped sum for NaCl ($h_{\rm max}=\sqrt{4/3}$), with a nearest neighbour distance of $1$, and $z_{1,2}=\pm 1$. The reference energy is derived from the NaCl Madelung energy ($E_{\rm ref}=M_{\rm NaCl}$) taken from Ref. \onlinecite{mamode2016computation} to 27 significant figures. The black line indicates an optimal path to convergence, and provides a relationship between $R_c$ and $R_c$: $R_c=3R_d^2/h_{\rm max}$.}\label{figure3}
\end{figure}

Compounds, for example the  SiO$_2$ and Al$_2$SiO$_5$ tested, can be considered as being comprised of two or more, subsystems of charge.  For SiO$_2$ the sub-lattices consist of the $+4$(Si) and $+6$(O) charges. The electrostatic energy for each may be evaluated separately within the current scheme. In the case of, for example, a defect in a large supercell, this will be computationally advantageous, since the different length scales associated with the defect, and the bulk lattice, will lead to an appropriate and computationally advantageous $R_c$ for each subsystem. Furthermore, the charges of the different subsystems may be of opposite signs. In this way the Madelung energy of ionic crystals can be evaluated. In Fig. \ref{figure3} the precision that can be achieved by the current approach is demonstrated through the calculation of the Madelung energy for the NaCl structure ($M_{\rm NaCl}$) and a range of $R_c$ and $R_d$. Reference values for $M_{\rm NaCl}$ are available to great precision,\cite{mamode2016computation} and the current scheme rapidly approaches this benchmark as damping is reduced and the cutoff sphere expanded.

\section{Derivatives}

The forces (derivatives of the energy with respect to the ionic positions) and stresses (derivatives with respect to lattice vectors) due to the pair interaction term in Eqn.\ref{dmpcut} can be evaluated in the normal way. Since the correction term, $\Delta E_i$, depends on the volume of the unit cell (through the density, $\rho$), there are additional contributions to the stress (but not the forces). This derivative of $\Delta E_i$ with respect to the volume of the unit cell (through the density $\rho$, and recalling that $R_a$ depends on $\rho$) is given by:
\begin{equation}
\begin{aligned}
\frac{\partial \Delta E_i}{\partial \Omega}=\frac{\pi}{3\Omega}Z_i\rho R_a^2\erfc(R_a/R_d)+\frac{\pi}{2\Omega}Z_i\rho R_d^2\erf(R_a/R_d)-\frac{\sqrt{\pi}}{\Omega}Z_i\rho R_aR_de^{-R_a^2/R_d^2}.
\end{aligned}
\end{equation}
The derivative of the total electrostatic energy with respect to the lattice vector coefficients $L_{\alpha\beta}$ is obtained by summing $\frac{\partial \Delta E_i}{\partial \Omega}$ over the ions, $i$, and multiplying the result by the volume $\Omega$ times the matrix of the reciprocal lattice vector coefficients.

\section{Relationship to Wolf's method}

\begin{figure}[]
\centering
\includegraphics[width=0.95\textwidth]{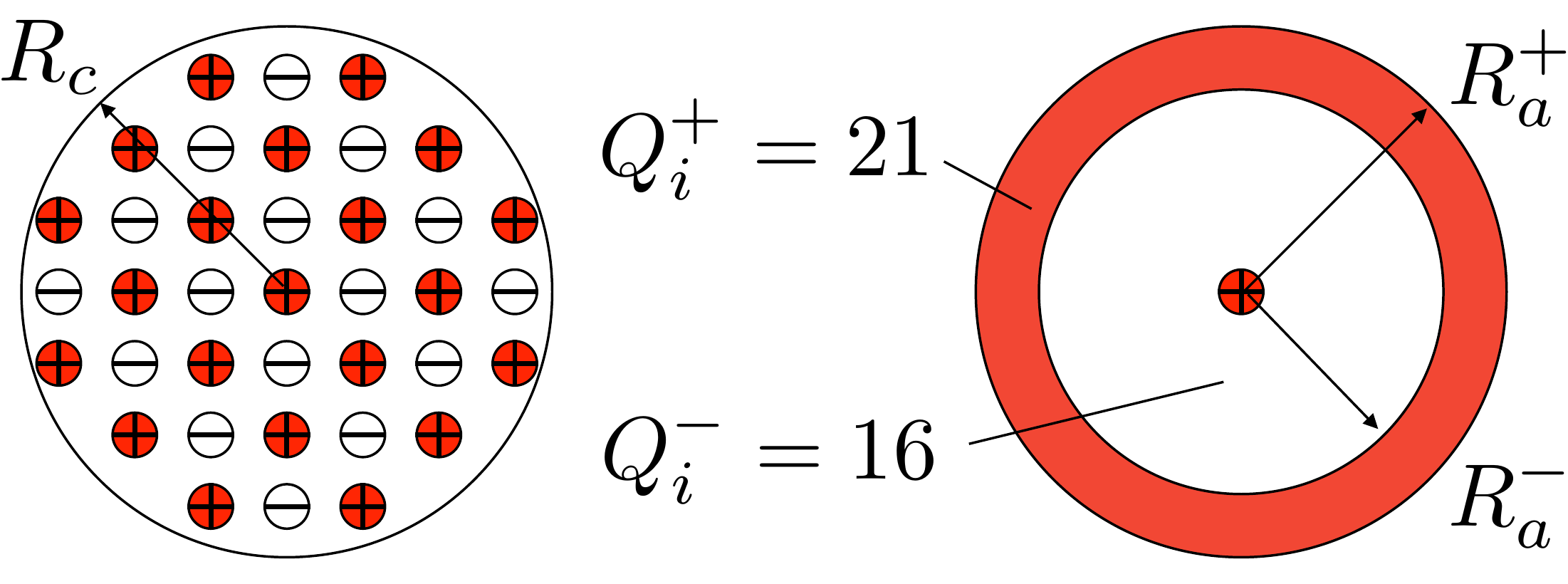}
\caption{Illustration of the relationship to Wolf's scheme.\cite{wolf1999exact} This spherically cutoff portion of a 2D ionic lattice contains $21$ positive ions, and $16$ negative ions. Following the current scheme, the adaptive radius for the positive ions is larger than that for the negative ions. In the region that the compensating spheres overlap, the net compensating charge density is precisely zero. All of the imbalance in the charge is distributed over a shell between $R_a^-$ and $R_a^+$. In Wolf's scheme, the compensating charge would be placed on a the surface of a sphere at exactly $R_c$.}\label{figure4}
\end{figure}

The application of the current scheme to the evaluation of the Madelung energy of ionic crystals allows direct comparison to Wolf's scheme. As illustrated in Fig. \ref{figure4}, the relationship between the two methods is particularly clear for the undamped situation. In the current scheme, for an overall charge neutral system, the positive and negative subsystems are individually neutralised by uniform densities of equal magnitude, but opposite signs. Since, in general, the total positive and negative charge enclosed by the cutoff sphere will not be equal, neither will the adaptive radii $R_a^+$ and $R_a^-$ for the positive and negative charge subsystems, respectively. As a result, a charge equal (and opposite in sign) to the difference between the number of positive and negative charges will be uniformly spread within a shell of inner radius $\min(R_a^+,R_a^-)$ and outer radius $\max(R_a^+,R_a^-)$. In the Wolf scheme this neutralising charge is placed on the surface of a sphere at precisely $R_c$. As the cutoff sphere expands, the two schemes approach each other, but lead to significantly different results for smaller $R_c$.

\section{Discussion}

The current scheme is more straightforward to implement than Ewald's method. There is no computationally costly reciprocal space summation to be performed (it is effectively replaced by the analytic correction term). Otherwise, the real space summation is identical to that in the Ewald scheme. As a result, any existing Ewald routine may be readily adapted to the new scheme. A direct implementation of the Ewald scheme leads to an O($N^2$) scaling, and optimised methods\cite{rajagopal1994optimized} scale from O($N^{3/2}$)\cite{fincham1994optimisation} to O($N\ln N$). \cite{darden1993particle} For a fixed $R_c$, the current method exhibits O($N$) scaling, which makes it of particular relevance to linear scaling density functional methods. Wolf's scheme has been found to be around a factor of 5 times faster than Ewald summation for charge neutral systems.\cite{chen2010atomic} Because the current scheme is closely related to Wolf's scheme it is expected to offer similar computational advantages, but for a wider class of systems. The numerical data presented here are calculated using quadruple precision, which is not straightforward to achieve for efficient Ewald implementations due to their dependence of optimised fast Fourier transform libraries, which are not typically available for arbitrary precision.

Many electronic structure methods perform a large number of electronic iterations for each ionic configuration, and the relative computational effort expended on the electrostatic summations is small (but cumulatively significant, given the large portion of global high performance computing dedicated to such calculations). Car-Parrinello molecular dynamics\cite{car1985unified} requires the more frequent reevaluation of the electrostatic summations, and so the relative computational effort expended on the electrostatic summations is greater. Attempts to accelerate the electronic structure updates will progressively reveal the cost of the electrostatic summations. This might be expected to be most significant for the so-called orbital free methods.\cite{wang2002orbital} In the case of a fixed unit cell, and fixed uniform charge density, the only dependence of Eqn. \ref{DFT} on the ionic positions is through the $E_{\rm NN}[\{{\bf r}_i\}]$ term. It can be seen that electrostatic summation in a uniform compensating background is in fact a simple variety of orbital free density functional method, and accounts for the entire computational cost.

Being a cutoff based method the computed forces are not strictly continuous as an ion moves across a sphere boundary. However, given the rapid convergence of the scheme this is not expected to cause significant problems in density functional applications. Should they arise, for a fixed $R_c$ the damping may be slightly increased, or for a fixed damping, $R_c$ may be increased to eliminate the discontinuity at the sphere boundary. Force shifting approaches might also be considered, as they have been in relation to the Wolf scheme, \cite{wolf1999exact,fennell2006ewald} along with discussion concerning the performance of non-Ewald methods for inhomogeneous systems.\cite{takahashi2011cutoff,wirnsberger2016non,elvira2014damped}

\section{Conclusion}

A scheme has been presented for the evaluation of the electrostatic energy for an extended collection of point charges. Crucially, it is applicable to the case that the overall system is not charge neutral -- with the charge imbalance being neutralised by carefully chosen spheres of uniform compensating charge. Density functional\cite{hohenberg1964inhomogeneous,kohn1965self} total energy\cite{martin2004electronic,payne1992iterative,wang2002orbital} methods require the evaluation of such quantities, as do other electronic structure methods, such as Quantum Monte Carlo based techniques.\cite{foulkes2001quantum} The scheme allows for the individual evaluation of contributions from subsystems of the point charges. In this way a neutral system can be treated, the result being closely related to Wolf's method. Damping the contribution from the point charges leads to a scheme that converges rapidly with the cutoff sphere radius, and a relationship between the cutoff radius and a suitable damping parameter is provided.

The straightforward physical motivation of this scheme, its algorithmic simplicity and the high accuracy and computational efficiency that can be achieved, suggests that it provides an attractive alternative to Ewald's scheme for modern and future electronic structure implementations.

\begin{acknowledgments}
CJP is supported by the Royal Society through a Royal Society Wolfson Research Merit award and the EPSRC through grants EP/P022596/1 and EP/J010863/2, and thanks Nigel Cooper, Matthew Foulkes, Peter Wirnsberger, Matt Probert and Daan Frenkel for their comments on the manuscript.

\end{acknowledgments}

%

\end{document}